%% file: root.tex
\documentclass[letterpaper, 10 pt, conference]{ieeeconf}  

\IEEEoverridecommandlockouts                              

\usepackage{graphics} 
\usepackage{epsfig} 
\usepackage{mathptmx} 
\usepackage{times} 
\usepackage{amsmath} 
\usepackage{amssymb}  
\usepackage{mathtools}  
\usepackage{multicol} 
\usepackage{hyperref} 
\usepackage{siunitx} 
\usepackage{subcaption} 

\usepackage{color,soul} 
\usepackage{marginnote}

\newcommand*{\xemspace}{\hspace{0.5em}}

\DeclareMathOperator{\sat}{sat}
\DeclareMathOperator*{\argmin}{arg\,min}
\DeclareSIUnit \var { VAr }
\newenvironment{proofref}[1]{\emph{Proof of Theorem~#1:}}{\hfill$\blacksquare$}

\title{\LARGE \bf
Optimal Control of Grid-Interfacing Inverters with Current Magnitude Limits
}

\author{Trager Joswig-Jones and Baosen Zhang
\thanks{Trager Joswig-Jones and Baosen Zhang are with the Department of Electrical and
        Computer Engineering, University of Washington Seattle, WA 98195, USA
        \{joswitra, zhangbao\}@uw.edu}
\thanks{The authors are partially supported by NSF grant ECCS-2153937.}%
}

\begin{document}

\maketitle
\thispagestyle{empty}
\pagestyle{empty}



\begin{abstract}
Grid-interfacing inverters act as the interface between renewable resources and the electric grid, and have the potential to offer fast and programmable responses compared to synchronous generators. With this flexibility there has been significant research efforts into determining the best way to control these inverters. An important nonlinear constraint in inverter control is a limit on the magnitude of the current, stemming from the need to protect  semiconductor devices. Existing approaches either simply saturate a controller that is designed for unconstrained systems, or assume small perturbations and linearize a saturated system. These approaches can lead to stability issues or limit the control actions to be too conservative.

In this paper, we directly focus on a nonlinear system that explicitly accounts for the saturation of the current magnitude. We use a Lyapunov stability approach to determine a stability condition for the system, guaranteeing that a class of controllers would be stabilizing if they satisfy a simple semidefinite programming condition. With this condition we fit a linear-feedback controller by sampling the output of (offline) model predictive control problems. This learned controller has improved performances with existing designs.


\end{abstract}

\section{INTRODUCTION}
\input{intro.tex}

\section{SYSTEM MODEL}
\input{model.tex}

\section{LYAPUNOV STABILITY}
\input{stability.tex}

\section{CONTROLLER DEVELOPMENT}
\input{controller.tex}

\section{SIMULATION RESULTS} \label{sec:simulation}
\input{simulation.tex}

\section{CONCLUSIONS}
\input{conclusion.tex}

%
\section*{APPENDIX}
\input{appendix.tex}


\bibliographystyle{IEEEtran}
\bibliography{Reference}

\end{document}

%% file: intro.tex
Electric power systems around the world are undergoing a dramatic transformation towards replacing conventional synchronous generation with renewable resources. Many of these resources, including solar photovoltaic, wind, storage, and electric vehicles are connected to the grid through power electronic inverters. In recent years, in addition to converting between DC and AC power, inverters are increasingly used to support the stability of the grid~\cite{matevosyan2019grid,guo2019performance}. 

Inverters typically offer grid support by changing their active/reactive power setpoints in response to changes in measured changes in frequency or voltage. A number of different control strategies have been proposed in the literature, including droop-control~\cite{schiffer2014conditions}, virtual oscillators and virtual synchronous machines~\cite{driesen2008virtual,dhople2013virtual}, neural network-based controllers~\cite{cui2022reinforcement} and others. Another line of research is to use inverters to explicitly shape the frequency response of the system, typically seeking to transform a second order system to a first order one~\cite{jiang2020dynamic,jiang2021grid}. 

In most of these works, the inverter is assumed to be ideal. That is, it can implement arbitrary active and reactive power (or voltage and current) commands. In practice, although inverters act much faster than conventional synchronous generators, they are also more limited in their actions. A key constraint for inverters is their current limit. A current limiter is an element that addresses over-currents that may appear during faults and voltage fluctuations, and damage sensitive semiconductor devices in inverters~\cite{qoria2020current}. Therefore, the output currents of inverters would saturate even if the command from other control loops calls for higher current~\cite{bottrell2013comparison}.

This type of saturation is nontrivial to analyze. In particular, currents are vectors in $\mathbb{R}^2$ (in the dq frame) and the saturation is on their magnitudes. Hence the tools for sector-constraints developed for Lur'e systems~\cite{park1997revisited} do not readily apply. Currently, current saturation is either ignored or analyzed in linearized~\cite{osorio2019robust} and reduced model systems~\cite{ajala2022model}. 

In this paper, we directly work with the nonlinear system and explicitly account for current magnitude saturation to design good performing controllers. In particular, we consider an inverter connected to an infinite bus and derive a condition on when a feedback controller stabilizes the system. This condition turns out to be quite geometrically intuitive and leads to a linear matrix inequality (LMI). Using this LMI, we learn the optimal controller by sampling a set of model predictive control (MPC) solutions. We show that this learned controller significantly outperforms existing controllers. 


%% file: model.tex
\subsection{Model}
\begin{figure}[ht]
    \centering
    \includegraphics[width=1.0\columnwidth, trim={0 0 0 0},clip]{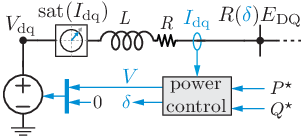}
    \caption{Simplified inverter system model under study.}
    \label{fig:simplified_model}
\end{figure}
%
In this paper, we consider a simplified model of a three-phase inverter connected to an infinite bus via an \emph{LCL} filter. The system operates in balanced three phase and uses a direct-quadrature (dq) reference frame with reference to the inverter voltage angle to describe all the rotating physical quantities~\cite{yazdani_vsc_book}. 
Consequently, all these quantities are vectors in $\mathbb{R}^2$. Our goal is to design a feedback controller such that the system tracks a constant power setpoint, $(P^*,Q^*)$, at the grid side. The dynamics of the system comes from the \emph{RL} filter element in the circuit, and we think of the current $I_\mathrm{dq}=(I_\mathrm{d},I_\mathrm{q})$ as the state of the system. We model the inverter itself as a voltage source converter, and treat its output, $V_\mathrm{dq}=(V_\mathrm{d},V_\mathrm{q})$ as the actuator output.
We take the the magnitude of the inverter voltage, $V$, and the angle difference between the inverter and grid voltages, $\delta$, to be the control outputs and choose the d-axis to align with the inverter voltage such that $V_d = V$ and $V_q = 0$\footnote{With a small-angle assumption this is equivalent to taking $V_\mathrm{dq}$ to be the control outputs.}. For the simplified model, we assume that the angle of the grid-voltage is known, however in the full-order model this is a measured value using a phase-lock loop.
We assume the grid side voltage $E$ has a constant magnitude such that $E_\mathrm{DQ} = (\sqrt{2} V_{\mathrm{nom}}, 0)$, where $V_{\mathrm{nom}}$ is the nominal phase-neutral voltage of the grid
\footnote{The assumption that $E_\mathrm{q}=0$ can be made without loss of generality, but the assumption $E_\mathrm{d}$ is constant is more material. An important future question is to allow time variation in $E_\mathrm{d}$.}. 
In the inverter reference frame, this is $E_\mathrm{dq}=(E_\mathrm{d},E_\mathrm{q})=R(\delta)\cdot E_\mathrm{DQ}$, where $R(\cdot)$ is a counter-clockwise rotation matrix.

With the the above notations and assumptions, the active and reactive powers at the grid side are given by 
$$ P = \frac{3}{\sqrt{2}} V_\mathrm{nom} \cdot I_\mathrm{d}, \xemspace Q = -\frac{3}{ \sqrt{2}} V_\mathrm{nom} \cdot I_\mathrm{q}.$$
Therefore, tracking a given $P^*,Q^*$ is equivalent to tracking some $I_\mathrm{d}^*$ and $I_\mathrm{q}^*$. We further make the following small signal assumptions. The first is that the angle difference between the inverter voltage and the grid voltage is small. Namely, let $\delta$ be the angle difference, and we use the approximations where $\sin(\delta)\approx \delta$ and $\cos(\delta)\approx 1$. The second assumption we make is that both the inverter and grid reference frames rotates at the same frequency, $\omega_{\mathrm{nom}}$. These two assumptions allows us to eliminate the nonlinear cross couplings in the \emph{RL} filter dynamics. 

For more details on how these simplification can be derived from a full-order inverter model with current and voltage control loops, please see Appendix~\ref{app:full_order}. With these assumptions, the current dynamics are linear in the voltages. 
The reason we make these assumptions is to isolate the challenge introduced by the nonlinear saturation block, which we describe next. 

\subsection{Current Saturation}
To protect internal devices, the inverter's output current cannot exceed a preset limit. Different than typical saturation limits in the literature, which limits each component of control input or the state, the limit in our case is on the the magnitude of the current vector. More precisely, we define
\begin{equation} \label{eq:magnitude_limiter_max}
\mathrm{sat}(z) = 
\begin{cases}
z & \text{if } \|z\|_2 \leq 1 \\
\frac{1}{\|z\|_2} \cdot z & \text{if } \|z\|_2 > 1
\end{cases} 
=  \frac{1}{\max{(\|z\|_2, 1)}} \cdot z. 
\end{equation}
This saturation on the magnitude\footnote{Without loss of generality, we can take the value of the saturated magnitude to be 1 when developing these theorems.} of the currents is not a sector inequality, therefore some of the results from Lur'e problems do not directly apply. The main theoretical result of this paper is that a simple geometric argument allows us to characterize the stability of a linear system with magnitude saturation. 


\subsection{System Model}
Due to space constraints, we directly state the discrete time dynamical system model. It comes from discretizing the differential equations governing the current through the inductor using the forward Euler method, with a saturation limit on the current:
\begin{equation} \label{eqn:system}
x_{t+1}=\sat\{A x_t +B u_t \}, 
\end{equation}
where 
\begin{align*}
    A &= I - \Delta t \begin{bmatrix}- \frac{R}{L} & \omega_\mathrm{nom}\\- \omega_\mathrm{nom} & - \frac{R}{L}\end{bmatrix}, \; \; B = \Delta t \begin{bmatrix}\frac{\sqrt{2}}{L} & 0\\0 & \frac{\sqrt{2} E}{L}\end{bmatrix} \\
     x_t &= \begin{bmatrix}I_{\mathrm{d}}{\left(t \right)}\\I_{\mathrm{q}}{\left(t \right)}\end{bmatrix}, \;\; u_t = \begin{bmatrix}V\\ \delta \end{bmatrix},
\end{align*}
where $I$ is the identity matrix and $\Delta t$ is a sufficiently small discretization timestep. 

Our goal is to design an optimal linear feedback controller such that the system in \eqref{eqn:system} is asymptotically stable around a setpoint $x^*$.

%% file: stability.tex
We assume that the tracking problem is feasible, that is, the reference $x^*$ satisfies $||x^{*}||_2 < 1$.
Let $\Delta x=x-x^*$ denote the shifted state with the reference at the origin. Note that the saturation is on $||\Delta x+x^*||_2$. In this paper, we study the class of linear feedback controllers with the gain matrix $K$, where $u = u^* -K (x - x^*)$. In this section we find a constraint on $K$, such that the saturated-state system is asymptotically stable and has a unique equilibrium. 

The system with shifted state $\Delta x$ is
\begin{subequations} \label{eqn:Dx}
\begin{align}
\Delta x_{t+1} & = A (\hat{x}_t - x^*) - B K (\hat{x}_t - x^*), \\
\hat{x}_t &= \mathrm{sat}(\Delta x_t + x^*),  \label{eqn:xhat}
\end{align}
\end{subequations}
where we introduce $\hat{x}_t$ as a variable that represents the saturated state at time $t$. We have the following result
\newtheorem{theorem}{Theorem}
\begin{theorem} \label{thm:circular-limiter-stability}
Consider the system in \eqref{eqn:Dx}. Suppose the reference and the initial starting point satisfy $||x^*||_2 <1$ and $||x_0||_2<1$, respectively. The system is asymptotically stable around $x^*$ if $K$ satisfies 
\begin{equation} \label{eqn:K}
(A - B K)^\top (A - B K) - I \prec 0.
\end{equation}
\end{theorem}

\begin{figure}
  \begin{subfigure}{0.48\columnwidth}
    \includegraphics[width=\linewidth, trim={30 30 30 30},clip]{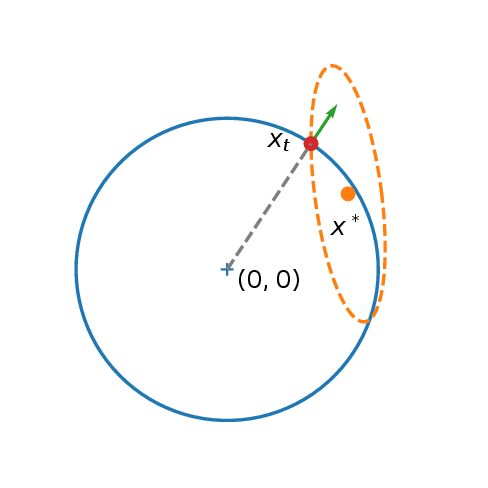}
      \caption{A case where $x_t$ is ''stuck'' and will remain stationary; never reaching $x^*$.
      \vspace{1em}
      }
      \label{fig:stuck_state}
  \end{subfigure}%
  \hspace*{\fill}   
  \begin{subfigure}{0.48\columnwidth}
    \includegraphics[width=\linewidth, trim={30 30 30 30},clip]{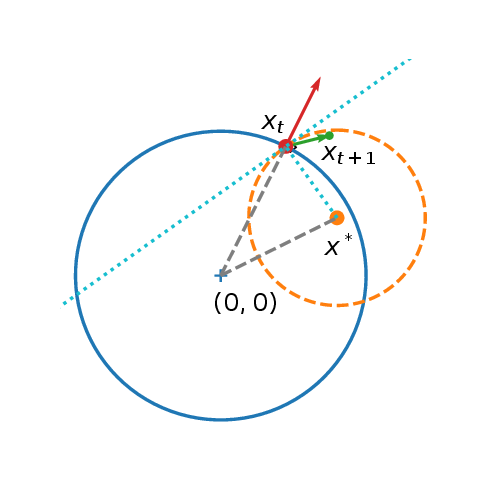}
    \caption{A case where the Lyapunov level set is circular and $x_{t+1} - x_t$ can not align with the saturation function normal $x_t$. 
    }
      \label{fig:circular_level_sets}
  \end{subfigure}%
  \vspace{5pt}

\caption{Illustrative figures showing the impact the shape of the Lyapunov level sets has on asymptotic stability. The saturation function boundary and the Lyapunov level sets are represented as ellipses with solid and dashed lines, respectively.} \label{fig:1}
\vspace{-10pt}  
\end{figure}

\vspace{3pt}  
The condition in \eqref{eqn:K} is the standard Lyapunov stability condition for linear systems where the $P$ matrix is chosen to be identity~\cite{linearHespanha2009dtLyapunov}. This means that the unsaturated system needs to be stable with a Lyapunov function that has circular level sets. The intuition behind this result is that if the Lyapunov function's level sets and the saturation function have the same shape, we can use triangle inequality to show that the trajectory of the system with saturation never gets ``stuck''. 
It is possible to find systems that are stable but do not have Lyapunov functions with circular level sets, where the trajectory of the saturated system can remain stationary at some point along the magnitude bound and not converge to the reference (see Section~\ref{sec:simulation}).
Fig.~\ref{fig:stuck_state} shows how without a circular level set the state dynamics can align with the normal of the circular saturation function, resulting in the state getting ''stuck'' in this position. 
Fig.~\ref{fig:circular_level_sets} visualizes how with a circular level set the normal of the saturation function circle and the dynamics step in the states cannot align.

\begin{proofref}{\ref{thm:circular-limiter-stability}}
We select the Lyapunov function $V(z) = z^\top z$. It is clear that $V(0)=0$. Then it suffices to show that 
$V(\Delta x_{t+1}) < V(\Delta x_{t}) \; \forall \; \Delta x_t  \in \mathbb{R}^2, \; \Delta x_t \neq 0$.

If $\| \Delta x_{t} + x^* \|_2 \leq 1$, then $\Delta x_{t+1} = A(\Delta x_{t}) + B u (\Delta x_{t})$. Then $V(\Delta x_{t+1}) < V(\Delta x_{t})$ by \eqref{eqn:K}.  
Therefore, we focus on when $\| \Delta x_{t} + x^* \|_2 > 1$, where $\hat{x}_t =  \frac{\Delta x_t + x^*}{\| \Delta x_t + x^* \|_2}$ from \eqref{eq:magnitude_limiter_max} and \eqref{eqn:xhat} such that we have the step dynamics
$$
\Delta x_{t+1} = (A - B K) \left( \frac{\Delta x_t + x^*}{\| \Delta x_t + x^* \|_2} - x^* \right).
$$
By \eqref{eqn:K}, we have
$$
V(\Delta x_{t+1}) < \left\| \frac{\Delta x_t + x^*}{\| \Delta x_t + x^* \|_2} - x^* \right\|^2_2
$$
Now we show that 
\begin{equation} \label{eqn:V_dec}
V(\Delta x_{t+1}) < \left\| \frac{\Delta x_t + x^*}{\| \Delta x_t + x^* \|_2} - x^* \right\|^2_2
<
V(\Delta x_{t}) = \| \Delta x_{t} \|^2_2. 
\end{equation}
%
%
Given the vector $x_t$, let $v$ to be a vector orthogonal to it. Then $\{x_t, v\}$ forms a basis in $\mathbb{R}^2$, and we can write $x^*$ as 
$$
x^* = c_1 x_t + c_2 v,
$$
where $c_1 = \frac{\langle x_t, x^* \rangle}{\|x_t\|_2^2}$ and the term $c_1 x_t$ represents the projection of $x^*$ onto $x_t$. Then 
\begin{equation} \label{eqn:Vt}
V(\Delta x_t)=\| x_{t} - x^* \|^2_2 = 
\left\|x_t - \frac{\langle x_t, x^* \rangle}{\|x_t\|_2^2} x_t \right\|^2_2 + \| c_2 v \|_2^2.    
\end{equation}

To show \eqref{eqn:V_dec}, we have
\begin{align*}
    V(\Delta x_{t+1}) & <\left\| \frac{\Delta x_t + x^*}{\| \Delta x_t + x^* \|_2} - x^* \right\|^2_2 \\
& \stackrel{(a)}{=}
\left\| \frac{1-\langle x_t,x^* \rangle}{\| \Delta x_t + x^* \|_2} x_t \right\|^2_2 + \| c_2 v \|_2^2 \\
& =\left(\frac{1}{\| x_t \|_2} - \frac{\langle x^*, x_t \rangle}{\| x_t \|_2^2} \right)^2 \| x_t \|_2^2 +  \| c_2 v \|_2^2 \\
& \stackrel{(b)}{<} \left(1 - \frac{\langle x^*, x_t \rangle}{\| x_t \|_2^2} \right)^2 \| x_t \|_2^2 +  \| c_2 v \|_2^2 \\
& = \left\|x_t - \frac{\langle x_t, x^* \rangle}{\|x_t\|_2^2} x_t \right\|^2_2 + \| c_2 v \|_2^2 \\
& \stackrel{(c)}{=} \| x_{t} - x^* \|^2_2, 
\end{align*}
where $(a)$ follows from $x_t$ and $v$ being orthogonal to each other and $(c)$ follows \eqref{eqn:Vt}. The step $(b)$ follows from the fact that  
$$
\frac{\langle x^*, x_t \rangle}{\| x_t \|_2^2} < \frac{1}{\| x_t \|_2} < 1.
$$
To see this, note that by assumption, we assume $x_t$ saturates, so $\| x_t \|_2^2 > 1$ and $ \frac{1}{\| x_t \|_2} < 1$. 
Next, we manipulate the inequality $\frac{\langle x^*, x_t \rangle}{\| x_t \|_2^2} < \frac{1}{\| x_t \|_2}$ by applying the Cauchy-Schwarz inequality ($| \langle x^*, x_t \rangle | \leq \| x^* \|_2 \cdot \| x_t \|_2$) to get
$$
\frac{\langle x^*, x_t \rangle}{\| x_t \|_2^2} 
\leq \frac{\| x^* \|_2 \cdot \| x_t \|_2}{\| x_t \|_2^2} 
= \frac{\| x^* \|_2}{\| x_t \|_2} 
< \frac{1}{\| x_t \|_2},
$$
which is true since we assume $\| x^* \|_2 < 1$. \end{proofref}

\subsection{Stability with unknown line parameters}
The resistance and inductance parameters---$R$ and $L$ defining the matrices $A$ and $B$---are the sum of the inverter side and grid side parameters. In practice, the inverter side $R$ and $L$ are known since they are part of the design parameters, but the grid side parameter depends on the exact connection setup and is often unknown. Therefore, it is useful to design $K$ to be robust to this uncertainty in the grid parameters. The next theorem provides a way to do this assuming that the grid-side line is inductive ($L_g > 0$)
\begin{theorem}\label{thm:R}
Suppose the gain matrix $K$ satisfies \eqref{eqn:K} for some given $A$ and $B$ matrices. Then $K$ satisfies \eqref{eqn:K} if the value of $R$ is increased, as long as $L > 0$ and the discretization steps $\Delta t$ are sufficiently small.
\end{theorem}
Therefore, if the parameters of the grid-side line inductance and resistance are unknown, then it suffices to design the controller assuming the grid side resistance is $0$ (the value of the inductance does not enter into stability). This leads to a suboptimal, but robust controller.

\begin{proofref}{\ref{thm:R}}
We have
$$
A =
I -
\Delta t \left[\begin{matrix} \frac{R}{L} & \omega_\mathrm{nom}\\- \omega_\mathrm{nom} & \frac{R}{L}\end{matrix}\right]
=
I -
\Delta t \hat{A}
$$

$$
B =
\Delta t \left[\begin{matrix}\frac{\sqrt{2}}{L} & 0\\0 & \frac{\sqrt{2} E}{L}\end{matrix}\right] = \Delta t \hat{B}
$$
where
$
\hat{A} = \left[\begin{matrix} \frac{R}{L} & \omega_\mathrm{nom}\\- \omega_\mathrm{nom} & \frac{R}{L}\end{matrix}\right] \mbox{ and } 
\hat{B} = \left[\begin{matrix}\frac{\sqrt{2}}{L} & 0\\0 & \frac{\sqrt{2} E}{L}\end{matrix}\right]. 
$
Suppose $K$ satisfies $(A - BK)^\top (A - BK) - I \prec 0.$ Expanding this inequality and neglecting the terms that contain $\Delta t^2$ (by the assumption that $\Delta t$ is sufficient small), we have 
$$
-\Delta t (\hat{A}^\top + K^\top \hat{B}^\top + \hat{A} + \hat{B} K) \prec 0,
$$
or
$$
\hat{A}^\top + K^\top \hat{B}^\top + \hat{A} + \hat{B} K \succ 0. 
$$
Writing the matrices out, we have 
$$
\left[\begin{matrix} 2\frac{R}{L} & 0 \\ 0 & 2\frac{R}{L}\end{matrix}\right]^\top
+ K^\top \left[\begin{matrix}\frac{\sqrt{2}}{L} & 0\\0 & \frac{\sqrt{2} E}{L}\end{matrix}\right]^\top
+ \left[\begin{matrix}\frac{\sqrt{2}}{L} & 0\\0 & \frac{\sqrt{2} E}{L}\end{matrix}\right] K \succ 0
$$
Since $L > 0$,  we obtain
$$
\left[\begin{matrix} 2 R & 0 \\ 0 & 2 R \end{matrix}\right]^\top
+ K^\top \left[\begin{matrix} \sqrt{2} & 0\\0 & \sqrt{2} E \end{matrix}\right]^\top
+ \left[\begin{matrix} \sqrt{2} & 0\\0 & \sqrt{2} E \end{matrix}\right] K \succ 0.
$$
Now we note that the left hand side expression is a monotonic in terms of $R$. Namely, if $K$ satisfies this inequality for $R_1$, then $K$ also satisfies this inequality for any $R_2 > R_1$.
\end{proofref}

%% file: controller.tex
In the last section, Theorems~\ref{thm:circular-limiter-stability} and~\ref{thm:R} provided stability conditions on the gain matrix $K$, but not optimality conditions for how to select a stabilizing $K$. A number of objective functions are used in practice to optimize the controller gain, including settling times, rate of change of current, overshoot, control efforts, and a combination of these. In general, there are no simple rules on how to find an optimal $K$. In this paper, we take a data-driven approach, where we first simulate a set of MPC controlled trajectories and then fit a $K$ that best approximates the MPC solutions. 
\subsection{Model Predictive Controller}

We formulate the optimal control problem for the discrete time system with saturation as the constrained optimization problem 
\begin{subequations} 
\label{eq:AC OPF}
\begin{align}
\label{eq:optimal-control-problem}
\displaystyle
u^{x_\mathrm{init}, x^*}_\text{ 
 mpc} := \argmin_{u_1, \dots, u_T} \sum_{i = 1}^{T} \quad & f(x_i, u_i, x^*)  \\
\textrm{s.t.} \quad &
\Delta x_{t+1} = A (x_t - x^*) + B u_t \\
\quad &
x_t = \mathrm{sat}(\Delta x_t + x^*)\\
\quad &
x_{0} = x_\mathrm{init}
\end{align}
\end{subequations}
where $x_\mathrm{init}$ and $x^*$ are the initial value of the states and the reference value for the controller, respectively, and $f(x_i, u_i, x^*)$ is a user selected objective function. We use this problem formulation to run a nonlinear model predictive controller. 

\subsection{Fitting a Static Controller}

Using the stability constraint from Theorem~\ref{thm:circular-limiter-stability} we formulate the linear regression problem:
\begin{subequations} 
\label{eq:fit-static-control-nonconvex}
\begin{align}
\displaystyle
\argmin_{K} \quad & \sum_{i\in U} \left( \sum_{t = 0}^{T_i - 1} \|-K (\Delta x_{t}^{i}) - u_{t}^{i}\|_2 \right) \\
\textrm{s.t.} \quad &
(A-BK)^\top(A-BK)-I\prec 0. \label{eq:nonlinear-constraint}
\end{align}
\end{subequations}
This problem is convex, since we can express \eqref{eq:nonlinear-constraint} as a linear matrix inequality using Schur complement as shown in~\cite{boyd_linear_1994_standard_problems}:
\begin{subequations}
\label{eq:fit-static-control-convex}
\begin{align}
\displaystyle
\argmin_{K} \quad & \sum_{i\in U} \left( \sum_{t = 0}^{T_i - 1} \|-K (\Delta x_{t}^{i}) - u_{t}^{i}\|_2^2 \right) \\
\textrm{s.t.} \quad &
\begin{bmatrix} 
I & (A-BK)^\top \\
(A-BK) & I 
\end{bmatrix} \succ 0. \label{eq:convex-constraint}
\end{align}
\end{subequations}
Here $U$ is a dataset containing sets of state values, $\Delta x^i \in \mathbb{R}^{n \times T_i}$ and associated MPC optimal input values, $u^i \in \mathbb{R}^{m \times T_i-1}$, where $n$ is the number of states, $m$ is the number of inputs, and $T_i$ is the length of simulation $i$

%% file: simulation.tex
To test our approach, we fit a controller to a series of MPC responses and we simulate the response of the system with the MPC-fit controller and a base-line controller for comparison. We select the base-line controller to be the linear–quadratic regulator (LQR) with
\begin{equation}
    K_\mathrm{base} = 
    \begin{bmatrix}
    1.206 & 0.0957 \\
    0.096 & 0.0671
    \end{bmatrix},
    \label{eq:Kbase}
\end{equation}
found with $Q = 
\begin{bmatrix}
1 & 0 \\
0 & 0.1
\end{bmatrix}$ 
and 
$R = 5\cdot B$. 
In the following we use the function $\mathrm{linspace}(a, b, n) \coloneqq a + \frac{{n-1}}{{i-1}}(b-a)$ to define a list of evenly spaced numbers over a specified interval, and the notation $(\cdot \times \cdot)$ to represent a Cartesian product of two sets.
 
We generate a set of MPC responses, 
$U \coloneqq \{x^{x_\mathrm{init}, x^*}_\text{  mpc}, u^{x_\mathrm{init}, x^*}_\text{  mpc} \mid (x_\mathrm{init}, x^*) \in X_\mathrm{init} \times X_\mathrm{ref} \}$.
$U$ contains the state trajectories and optimal MPC inputs for each $(x_\mathrm{init}, x^*)$ pair from a mesh of values in the polar grids, 
$X \coloneqq \{ r \cos{\theta + \frac{\pi}{4}} \mid (r, \theta) \in R \times \Theta \}$, 
where 
$R \coloneqq \mathrm{linspace}(0, I_\mathrm{max}, 3)$
and 
$\Theta \coloneqq \mathrm{linspace}(0, 2 \pi - \frac{2 \pi}{4}, 4)$.
$I_\mathrm{max}$ is the current magnitude limit of the inverter.
%

We use GurobiPy \cite{gurobi-2023} to solve a rolling time horizon version of \eqref{eq:optimal-control-problem} with an LQR objective $\sum_{i = 1}^{T} \quad  (x_i - x^*)^\top Q (x_i - x^*) + u_{i}^\top R u_{i}$ where 
$Q$ and $R$ are the same matrices used to design the LQR base-line controller. We choose a discretization step size of $\Delta t = 10~\unit{\micro\s}$ and use parameter values that are modified from those used in \cite{ajala2022model} for the inverter and \emph{LCL} filter system. The full list of parameter values 
can be found in Table.~\ref{table:parameters} in the Appendix. 
The optimization horizon used is $5$ periods and each simulation runs until the states reach a steady-state value; $\|x_t - x_{t-1}\|_2 < 10^{-5}$ for $10$ consecutive periods. 

We then fit a linear feedback controller $K_\mathrm{fit}$ by solving \eqref{eq:fit-static-control-convex} with our dataset $U$ using CVXPY \cite{diamond2016cvxpy}. We find the optimal fit matrix to be
\begin{equation}
    K_\mathrm{fit} = 
    \begin{bmatrix}
    0.608 & 0.027 \\
    0.012 & 0.026
    \end{bmatrix}.
    \label{eq:Kfit}
\end{equation}
%
%
We then test this MPC-fit controller and base-line controller on the simplified system by running simulations with all $(x_\mathrm{init}, x^*)$ in the set $X$. The results of these tests are detailed in the next section. The code used to generate these results is available at \href{https://github.com/TragerJoswig-Jones/Current-Magnitude-Limit-Inverter-Control}{https://github.com/TragerJoswig-Jones/Current-Magnitude-Limited-Inverter-Control}.

\subsection{Results}
The average cost of control for the MPC-fit and base-line controllers were 59.2 and 115.1, respectively. While the fit controller outperformed the base-line controller in terms of the average cost, the base-line controller showed lower costs for a majority of the test cases. However, for some test cases (22 to be exact) the base-line controller got stuck along the boundary and did not converge to the reference value.

\begin{figure}
      \centering
      \includegraphics[width=\columnwidth, trim={0 8 0 8},clip]{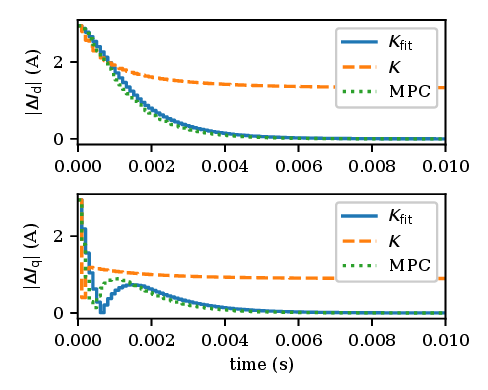}
      \caption{$| \Delta I_\mathrm{dq} |$ trajectories for $K, K_\mathrm{fit}$, and MPC controllers with $x_0 = (0.00~\unit{\ampere}, 0.00~\unit{\ampere})$, $x^* = (2.95~\unit{\ampere}, 2.95~\unit{\ampere})$}
      \label{fig:K-vs-Kfit-time-vs-idq}
    \end{figure}

\begin{figure}
      \centering      \includegraphics[width=\columnwidth, trim={7 8 0 8},clip]{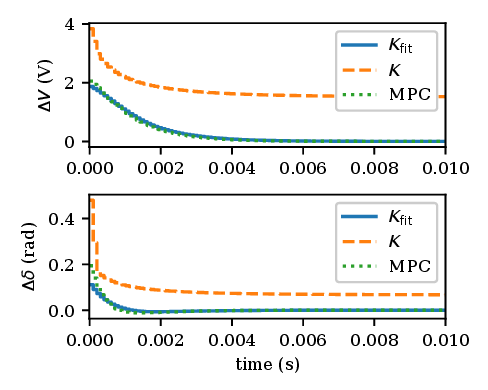}
      \caption{$\Delta V, \Delta \delta$ inputs for $K, K_\mathrm{fit}$, and MPC controllers with $x_0 = (0.00~\unit{\ampere}, 0.00~\unit{\ampere})$, $x^* = (2.95~\unit{\ampere}, 2.95~\unit{\ampere})$}
      \label{fig:K-vs-Kfit-time-vs-Vdelta}
    \vspace{-10pt}  
    \end{figure}

Fig.~\ref{fig:K-vs-Kfit-time-vs-idq} shows the trajectory of the absolute value of the state errors for the MPC-fit controlled and base-line controlled systems for a selected test cases. In this case the base-line controller gets stuck at the circular magnitude bound and does not converge to the reference value. We note that for this selection of $Q$ and $R$ the LQR controller, $K_\mathrm{base}$, does not satisfy Theorem~\ref{thm:circular-limiter-stability}. In this case, the MPC-fit controller converges to the reference value along a similar trajectory to the MPC controlled system.
The control inputs of the MPC-fit controlled system and the MPC controlled system align well, as seen in Fig.~\ref{fig:K-vs-Kfit-time-vs-Vdelta}, further demonstrating that the static linear feedback controller can adequately reproduce the optimal inputs from a rolling horizon MPC controller for this system and choice of LQR objective function.

%% file: conclusion.tex
In this paper, we presented a simple semidefinite programming condition for a simplified model of the nonlinear inverter system with current magnitude saturation that guarantees that a class of controllers would be stabilizing. With this condition we fit a linear-feedback controller to sampled data from nonlinear MPC problems that are run offline. This approach can be used with any objective function for the MPC problems. We found that the fit linear feedback controller can imitate well the MPC controller with an LQR objective guaranteeing stability and convergence to the reference in spite of the current magnitude saturation. 
Future work includes considering the impact of disturbances in the grid voltage, investigating other objective functions, and considering the cost of actuation required to maintain the current magnitude saturation bound.
%

%% file: appendix.tex
\subsection{Full-order system model} \label{app:full_order} 

The full-order system has the dynamics 
\begin{align}
\dot{x}=\begin{bmatrix}
\dot \delta_\mathrm{pll}\\
\dot \Pi_\mathrm{pll}\\
\dot \Phi_\mathrm{d}\\
\dot \Phi_\mathrm{q}\\
\dot \Gamma_\mathrm{d}\\
\dot \Gamma_\mathrm{q} \vspace{0.25em}\\
\dot I_{i\mathrm{d}}\vspace{0.25em}\\
\vspace{0.25em}\dot I_{i\mathrm{q}}\vspace{0.25em}\\
\vspace{0.25em}\dot W_{\mathrm{d}}\vspace{0.25em}\\
\vspace{0.25em}\dot W_{\mathrm{q}}\vspace{0.25em}\\
\vspace{0.25em}\dot I_{g\mathrm{d}}\vspace{0.25em}\\
\vspace{0.25em}\dot I_{g\mathrm{q}}
\end{bmatrix} = 
\begin{bmatrix}k_{i_{pll}} \Pi_\mathrm{pll} + k_{p_{pll}} W_{\mathrm{q}}\\
W_{\mathrm{q}}\\
\sqrt{2} V \cos{\left(\delta_{i} \right)} - W_{\mathrm{d}}\\
\sqrt{2} V \sin{\left(\delta_{i} \right)} - W_{\mathrm{q}}\\- I_{i\mathrm{d}} + \mathrm{sat}{\left(I_{i\mathrm{d}}^* \right)}\\- I_{i\mathrm{q}} + \mathrm{sat}{\left(I_{i\mathrm{q}}^* \right)}\\
\omega_\mathrm{ref} I_{i\mathrm{q}} - \frac{R_{i} I_{i\mathrm{d}}}{L_{i}} + \frac{V_{\mathrm{d}} - W_{\mathrm{d}}}{L_{i}}\\- \omega_\mathrm{ref} I_{i\mathrm{d}} - \frac{R_{i} I_{i\mathrm{q}}}{L_{i}} + \frac{V_{\mathrm{q}} - W_{\mathrm{q}}}{L_{i}}\\\omega_\mathrm{ref} W_{\mathrm{q}} + \frac{- I_{g\mathrm{d}} + I_{i\mathrm{d}}}{C}\\- \omega_\mathrm{ref} W_{\mathrm{d}} + \frac{- I_{g\mathrm{q}} + I_{i\mathrm{q}}}{C}\\
\omega_\mathrm{ref} I_{g\mathrm{q}} - \frac{R_{g} I_{g\mathrm{d}}}{L_{g}} + \frac{- \sqrt{2} E \cos{\left(\delta_\mathrm{pll} \right)} + W_{\mathrm{d}}}{L_{g}}\\
- \omega_\mathrm{ref} I_{g\mathrm{d}} - \frac{R_{g} I_{g\mathrm{q}}}{L_{g}} + \frac{\sqrt{2} E \sin{\left(\delta_\mathrm{pll} \right)} + W_{\mathrm{q}}}{L_{g}}
\end{bmatrix}
\end{align}
where  
\begin{equation}
u=\left[\begin{matrix}V\\\delta_{i}\end{matrix}\right] = -K \left[\begin{matrix}
- I_\mathrm{d}^{*} + I_{g\mathrm{d}}\\
- I_\mathrm{q}^{*} + I_{g\mathrm{q}}
\end{matrix}\right]
+
\begin{bmatrix}
    V^* \\
    \delta^*
\end{bmatrix}
\label{eq:full-order-input}
\end{equation}
and
$I_{i\mathrm{d}}^*$, $I_{i\mathrm{q}}^*$, $V_\mathrm{d}$, $V_\mathrm{q}$, $\omega_\mathrm{ref}$ are defined as
%
\begin{equation}
\resizebox{1.0\hsize}{!}{$  
\renewcommand*{\arraystretch}{1.1}  
\left[\begin{matrix}
I_{i\mathrm{d}}^*\\
I_{i\mathrm{q}}^*\\
V_{\mathrm{d}}\\
V_{\mathrm{q}}\\
\omega_\mathrm{ref}
\end{matrix}\right] 
= 
\left[\begin{matrix}
- C \omega_\mathrm{nom} W_{\mathrm{q}} + k_{iv} \Phi_\mathrm{d} + k_{pv} (\sqrt{2} V \cos{\left(\delta_{i} \right)} - W_{\mathrm{d}}) + I_{g\mathrm{d}}\\ 
C \omega_\mathrm{nom} W_{\mathrm{d}} + k_{iv} \Phi_\mathrm{q} + k_{pv} (\sqrt{2} V \sin{\left(\delta_{i} \right)} - W_{\mathrm{q}}) + I_{g\mathrm{q}}\\ 
- L_{i} \omega_\mathrm{nom} I_{i\mathrm{q}} + k_{ii} \Gamma_\mathrm{d} + k_{pi} 
I_{i\mathrm{d}}^* - I_{i\mathrm{d}}  
+ W_{\mathrm{d}}\\ 
L_{i} \omega_\mathrm{nom} I_{i\mathrm{d}} + k_{ii} \Gamma_\mathrm{q} + k_{pi} 
I_{i\mathrm{q}}^* - I_{i\mathrm{q}}  
+ W_{\mathrm{q}} \\ 
\dot{\delta}_\mathrm{pll}
\end{matrix}\right]
$}
\label{eq:full-order-algebraic-equations}
\end{equation}
%
%
where the states $x$ contains the phase-lock loop (PLL) angle difference from the grid voltage ($\delta_\mathrm{pll}$), the PLL integral term ($\Pi_\mathrm{pll}$), the voltage controller integral terms ($\Phi_\mathrm{dq}$), the current controller integral terms ($\Gamma_\mathrm{dq}$), the inverter-side inductor currents ($I_{i\mathrm{dq}}$), the capacitor voltages ($W_\mathrm{dq}$), and the grid-side inductor currents ($I_{g\mathrm{dq}}$). 
Lastly, we note that we can find current references from power references as
\begin{equation}
\begin{bmatrix}
    I_d^* \\
    I_q^* \\
\end{bmatrix}
=
\begin{bmatrix}
    V_\mathrm{d}^* & V_\mathrm{q}^* \\
    V_\mathrm{q}^* & -V_\mathrm{d}^* \\
\end{bmatrix}
\begin{bmatrix}
    P^* \\
    Q^* \\
\end{bmatrix},
\label{eq:PQtoIdq}
\end{equation}
if we solve for $(V_\mathrm{d}^*, V_\mathrm{q}^*) = (V^* \cos{\delta^*}, V^* \sin{\delta^*})$ from steady-state power flow equations with the stiff grid voltage and given power references \cite{grainger1994power}.
The dynamics of the inner control loops are from \cite{ajala2022model} and the dynamics of the synchronous reference frame phase-lock loop from \cite{nicolini_pll_2020}. 
The control parameter values are selected according to the methods used in these papers and can be found in our code repository.

To simplify the model we make the following assumptions. The voltage used in power calculations, does not significantly vary from the nominal value, such that the powers are linear in the currents. The angle difference between the inverter voltage and the grid voltage is relatively small, such that we can make the small-angle approximations $\sin{(\delta)} = \delta, \cos{(\delta)} = 1$. The reference frame rotates at a rate close to the nominal frequency, such that $\omega_\mathrm{ref} \approx \omega_\mathrm{nom}$. The inner voltage and current control loops act fast enough, such that we can ignore the dynamics of the capacitor.

\subsubsection{Comparison of simplified and full-order model}
\begin{figure}
      \centering
      \includegraphics[width=\columnwidth, trim={0 8 0 8},clip]{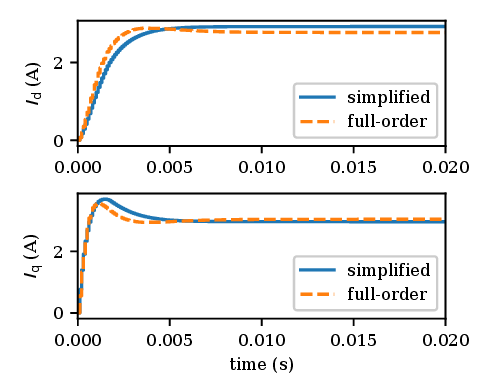}
      \caption{dq currents of the simplified and full-order systems for a step in $(P^*, Q^*)$ from $(0~\unit{\watt}, 0~\unit{\var})$ to $(775~\unit{\watt}, -775~\unit{\var})$}    
      \label{fig:Idq-vs-time-full-order-and-simplified-systems-Idq-ref-step}
      \vspace{5pt}  
    \end{figure}

We compare the response of the full-order and simplified systems with the MPC-fit controller by simulating the response to steps in $P^*, Q^*$. $I_d^*, I_q^*$ are found from $P^*, Q^*$ using \eqref{eq:PQtoIdq}.
%
%
%
We compare the grid-side currents $I_{g\mathrm{dq}}$ of the full-order system to the currents of the simplified model. Fig.~\ref{fig:Idq-vs-time-full-order-and-simplified-systems-Idq-ref-step} shows that both systems display similar trajectories and settling times. Discrepancies in the trajectories are likely due to the dynamics and steady-state response of the capacitor.
%
%
%
    \begin{figure}
      \centering      \includegraphics[width=\columnwidth, trim={0 8 0 8},clip]{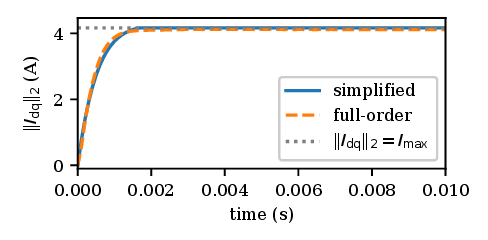}
      \caption{Current magnitude of the simplified and full-order systems for a step in $(P^*, Q^*)$ from $(0~\unit{\watt}, 0~\unit{\var})$ to $(775~\unit{\watt}, -775~\unit{\var})$}
      \label{fig:llIdqll-vs-time-full-order-and-simplified-systems-Idq-ref-step}
      \vspace*{-15pt}  
    \end{figure}

\subsection{Simulation parameters}
The parameter values used in section~\ref{sec:simulation} are listed in Table~\ref{table:parameters}. The full-order system parameter values are not included for brevity and can be found in the code repository. 
\begin{table}
\caption{Inverter \& \emph{LCL} filter simplified system parameters.}
\begin{center}
\begin{tabular}{c|c}
Parameter & Value \\
\hline
$V_\mathrm{nom}$    & 120~\unit{\volt} \\
\hline
$S_\mathrm{nom}$    & 1.5~\unit{\kilo\volt\ampere} \\
\hline
$I_\mathrm{nom}$    & 4.167~\unit{\ampere} \\
\hline
$I_\mathrm{max}$    & 4.167~\unit{\ampere} \\
\hline
\end{tabular}
\hspace{1em}
\begin{tabular}{c|c}
Parameter & Value \\
\hline
$E$   & 120~\unit{\volt} \\
\hline
$\omega_\mathrm{nom}$    & $2\pi60$~\unit{\radian\per\s} \\
\hline
$R$         & 1.3~\unit{\ohm} \\
\hline
$L$     & 3.5~\unit{\milli\henry} \\
\hline
\end{tabular}
\label{table:parameters}
\end{center}
\vspace*{-30pt}  
\end{table}
%